\documentclass[showpacs,aps,twocolumn]{revtex4}
\usepackage{graphicx}
\usepackage{amsfonts}
\usepackage{amsmath}
\usepackage{amssymb}
\usepackage{multirow}
\usepackage{braket}
\usepackage[usenames,dvipsnames]{color}

\begin{document}
\bibliographystyle{apsrev}

\title{Decomposition of conditional probability  for high-order symbolic Markov chains}

\author{S.~S.~Melnik and O.~V.~Usatenko}
\affiliation{A. Ya. Usikov Institute for Radiophysics and Electronics Ukrainian
Academy of Science, 12 Proskura Street, 61805 Kharkov, Ukraine}

\begin{abstract}
The main goal of the paper is to develop an estimate for the
conditional probability function of random stationary ergodic
symbolic sequences with elements belonging to a finite alphabet. We
elaborate a decomposition procedure for the conditional probability
function of sequences considered as the high-order Markov chains. We
represent the conditional probability function as the sum of
multi-linear memory function monomials of different orders (from
zero up to the chain order). This allows us to construct artificial
sequences by method of successive iterations taking into account at
each step of iterations increasingly more high correlations among
random elements. At weak correlations, the memory functions are
uniquely expressed in terms of the high-order symbolic correlation
functions. The proposed method fills up the gap between two
approaches: the likelihood estimation and the additive Markov
chains. The obtained results might be used for sequential
approximation of artificial neural networks training.
\end{abstract}

 \maketitle
%
\section{Introduction}

Systems with long-range interactions (and/or sequences with
long-range memory) and natural sequences with nontrivial information
content have been the focus of a large number of studies in
different fields of science for the past several decades.

Such random sequences are the subject of study of the algorithmic
(Kolmogorov-Solomonoff-Chaitin) complexity, information theory,
compressibility of digital data, statistical inference problem,
computability, data compression~\cite{Salomon}, natural language
processing~\cite{Mann}, artificial intelligence~\cite{Wiss} and have
many application aspects.

Random sequences with a \emph{finite number of state space} exist as
natural sequences (DNA or natural language texts) or arise as a
result of coarse-grained mapping of the evolution of the chaotic
dynamical system into a string of symbols~\cite{Eren,Lind}. The
items of sequence can also be phonemes, syllables, words or DNA's
base pairs according to the application.

A standard method of understanding and describing statistical
properties of a given random symbolic sequence of data requires the
estimation of the joint probability function of $L$-words
(subsequences of the length $L$) occurring. Reliable estimations for
the probability can be achieved only for small $L$ because the
number $m^L$ (where $m$ is the finite-alphabet length) of different
words of the length $L$ has to be much less than the total number of
words $m^L\ll M - L$ in the whole sequence of the length $M$.  This
is the crucial point because usually the correlation lengths of
natural sequences of interest are of the same order as the length of
sequence, whereas the last inequality can be fulfilled for the words
maximal lengths $L_{max} \lesssim 10$ only.

The main purpose and results of the paper can be formulated as
following. We elaborate a decomposition procedure for the
conditional probability function of sequences considered as the
high-order Markov chains. We represent the conditional probability
function as the sum of multi-linear monomials of different orders
(from zero up to the chain order).  At weak correlations, the memory
functions are uniquely expressed in terms of the high-order symbolic
correlation functions.

At last, we reveal close connection between our analytical Markov
chain approach and artificial neuron network models.

\section{Symbolic Markov chains}

Consider a semi-infinite random stationary ergodic sequence
$\mathbb{S}$  of symbols (letters, characters) $a_{i}$,
\begin{equation}
\label{ranseq}
 \mathbb{S}= a_{0}, a_{1},a_{2},...
\end{equation}
taken from the finite alphabet
\begin{equation}\label{alph}
 \mathcal{A}=\{\alpha^1,\alpha^2,...,\alpha^m\},\,\, a_{i}\in \mathcal{A},\,\, i \in
\mathbb{N}_{+} = \{0,1,2...\}.
\end{equation}

We use the notation $a_i$ to indicate a position $i$ of the symbol
$a$ in the chain and the unified notation $\alpha^k$ to stress the
value of the symbol $a\in \mathcal{A}$. We also use the personified
notation for the symbols $a$ of the same alphabet,
$\mathcal{A}=\{\alpha,\beta,...,\omega\}.$

We suppose that the symbolic sequence $\mathbb{S}$ is the
\textit{high-order Markov chain}. The sequence $\mathbb{S}$ is the
Markov chain if it possesses the following property: the probability
of symbol~$a_i$ to have a certain value $\alpha^k \in \mathcal{A}$
under the condition that {\emph{all}} previous symbols are fixed
depends only on $N$ previous symbols,
\begin{eqnarray}\label{def_mark}
&& P(a_i=\alpha^k|\ldots,a_{i-2},a_{i-1})\\[6pt]
&&=P(a_i=\alpha^k|a_{i-N},\ldots,a_{i-2},a_{i-1}).\nonumber
\end{eqnarray}

There are many other terms for such sequences. They are also
referred to as: \emph{categorical}~\cite{Hoss},
\textit{higher-order}~\cite{Raftery, Seifert}, \emph{the multi-} or
the $N$-\emph{step}~\cite{RewUAMM, UYa} Markov's chains. One of the
most important and interesting application of the symbolic sequences
is the probabilistic language model specializing in predicting the
next item in such a sequence by means of $N$ previous known symbols.
Here the Markov chain is known as the $N$-\emph{gram model}.

As a rule, the statistical properties of random sequences are
determined by correlation functions, however, in distinction from
the ordinary, numeric correlation functions,
\begin{eqnarray}\label{def_cor num}
\!\!\!\!\!\!\!\!&&C(r_1,r_2,\ldots,r_{k-1}) \\ [6pt]
\!\!\!\!\!\!\!&&\!\!\!\!\! = \overline{
[a_{0}\!-\!\overline{a})][a_{r_1}\,-\,\overline{a})] \ldots
[a_{r_1+\ldots+r_{k-1}}-\overline{a}]}, \nonumber
\end{eqnarray}
the \emph{symbolic correlation functions} of the $k$th order is
given by the following expression,
\begin{eqnarray}\label{def_cor1}
&&C_{\beta_1,...\beta_k}(r_1,r_2,\ldots,r_{k-1}) \\ [6pt]
=&&\overline{ [\delta(a_{0},\beta_1)-p_{\beta_1}]\ldots
[\delta(a_{r_1+\ldots+r_{k-1}},\beta_k)-p_{\beta_k}]}.\nonumber
\end{eqnarray}
The $\overline{overlines}$ mean a statistical average over an
ensemble of sequences. It can be replaced by the average along the
chain or by the arithmetic, Ces\`{a}ro's average for numerical
purposes. Note, that in some sense symbolic correlation
functions-matrices are more general construction than numeric
correlation functions. They can describe in more detail even numeric
sequences.
\subsection{Likelihood estimation}
If the sequence, the statistical properties of which we would like
to analyze, is given, then the conditional probability function
(CPF) of the $N$th order can be found by a standard method known
under the name of the likelihood estimation,
\begin{equation}\label{soglas}
P(a_i=\alpha|a_{i-N},\ldots,a_{i-1}) =\frac{
P(a_{i-N},\ldots,a_{i-1},\alpha) } { P(a_{i-N},\ldots,a_{i-1})},
\end{equation}
where $P(a_{i-N},\ldots,a_{i-1},\alpha)$ and
$P(a_{i-N},\ldots,a_{i-1})$ are the probabilities  of the
$(N+1)$-subsequence $a_{i-N},\ldots,a_{i-1},\alpha$ and
$N$-subsequence $a_{i-N},\ldots,a_{i-1}$ occurring, respectively.
Hereafter, we often drop the superscript $k$ from $\alpha^k$ to
simplify the notations.


\section{Formulation of the problem}
The conditional probability function completely determines \emph{all
statistical properties} of the random chain and the method of its
generation. Equation~(\ref{soglas}) says that the CPF is determined
if we know the probability of $(N+1)$-words occurring, the words
containing $(N+1)$ symbols without omissions among their numbers.
This instrument perfectly works if the random chain under study
satisfies the inequality
\begin{equation}\label{Lmax}
N<L_{max}\sim \frac{\ln M}{\ln m}.
\end{equation}
For $N>L_{max}$ this method represents adequately statistical
properties of the chain at $L<L_{max}$ only. The main idea of the
present paper is to elaborate a method, which would allow to use the
information contained in the ``perforated'', holed words such as,
e.g., $(a_{i}a_{i+3}), (a_{i}a_{i+5}), (a_{i}a_{i+N}),
(a_{i}a_{i+5}a_{i+8})$ and so on.

\subsection{Additive high-order Markov chains}

The unsatisfactory conclusion expressed by inequality
Eq.~(\ref{Lmax}), often referred to as the problem of the ``curse of
dimensionality", induced us to propose \cite{MUYaG,muya} an
alternative method based on the \emph{additive} high-order Markov
chain, where the conditional probability function takes on the
specific, simplified, ``linear form'' with respect to the random
variables $a_{i}$
\begin{eqnarray} \label{Dev_eq2}
&&P_{add}=P^{(1)}(a_i=\alpha|a_{i-N},\ldots,a_{i-2},a_{i-1})  \nonumber \\
\!&&\!= p_{\alpha}+\sum_{r=1}^N \! \sum_{\beta \in \mathcal{A}}\!\!
F_{\alpha \beta}(r)[\delta(a_{i-r},\beta)-p_{\beta}].
\end{eqnarray}
Here $p_{\beta}$ is the relative number of symbols $\beta$ in the
chain, or their probabilities of occurring,
\begin{equation}\label{a-av}
p_{\beta}=\overline{\delta(a_i, \beta)}.
\end{equation}
The Kronecker delta-symbol $\delta(.,.)$ plays the role of the
characteristic function of the random variable $a_i$ and converts
symbols to numbers. The additivity of the chain means that the
``previous'' symbols $a_{i-N},\ldots,a_{i-2},a_{i-1} \equiv
a_{i-N}^{i-1}$ exert an independent effect on the probability of the
``final``, generated symbol $a_i=\alpha$ occurring. The first term
in the right-hand side of Eq.~(\ref{Dev_eq2}) is responsible for the
correct reproduction of statistical properties of uncorrelated
sequences, the second one takes into account and correctly
reproduces correlation properties of the chain up to the second
order. The higher order correlation functions are not independent
anymore. We cannot control them and
reproduce correctly by means of the memory function $F(r)$.

There were suggested two methods for finding the \emph{memory
functions} $F_{\alpha \beta}(r)$ of a sequence with a known pair
correlation functions. The first one is a completely probabilistic
straightforward calculation analogous to that presented
in~\cite{MUYaG}. Its modification is used below while obtaining
Eqs.~(\ref{BilinCorr2}) and (\ref{BilinCorr3}). For any values of
$\alpha,\beta \in \mathcal{A}$ and $r \geqslant 1$ the relationship
between the correlation and memory functions was obtained,
\begin{equation}\label{Dev_eq4}
C_{\alpha \beta}(r)=\sum_{r'=1}^{N} \sum_{\gamma \in
\mathcal{A}}C_{\alpha \gamma}(r-r') F_{\beta \gamma}(r').
\end{equation}

Here the two-point (binary, pair) symbolic correlation function is
the particular case of definition \eqref{def_cor1},
\begin{equation}\label{Dev_eq1}
C_{\alpha \beta}(r)\!= \!\overline{\big[\delta(a_i,
\alpha)-p_{\alpha} \big]\!\big[ \delta(a_{i+r},
\beta)-p_{\beta}\big]}
.
\end{equation}

The second method~\cite{muya} for deriving Eq.~(\ref{Dev_eq4}) is
based on the minimization of the ``distance'' between the
conditional probability function, containing the sought-for memory
function, and the given sequence $\mathbb{S}$ of symbols with a
known correlation function,
\begin{equation}\label{Dev_eq40}
Dist = \overline{[\delta(a_{i},\alpha) -
P(a_i=\alpha|a_{i-N}^{i-1})]^2}.
\end{equation}

Let us note, that in the considered case the two-point quantities,
the memory and correlation functions, are not obliged to satisfy
strong inequality~\eqref{Lmax}. In other words the additive Markov
chain,~\eqref{Dev_eq2}, can describe and predict two-point
statistical properties of random sequences at distances longer than
that based on Eq.~\eqref{def_cor num}. At the same time, the model
of random sequences based on the likelihood estimation works more
satisfactory at short distances; see, e.g., the result of the DNA's
entropy estimation~\cite{MU,DNA16Mus}, where the discrepancy between
the two theories is evident. The next step to improve the prediction
quality of symbols in random chain, which we present here, is based
on the $k$-linear conditional probability function, which should
close the mentioned gap and, probably, explain a very astonishing
fact why binary correlations are so important and so long.

\section{Decomposition of the conditional probability function}
Equation~(\ref{Dev_eq2}) can be considered as an approximate model
expression simplifying the general form of the conditional
probability function. As a matter of fact the conditional
probability (\ref{soglas}) of the symbolic sequence of random
variables $a_i\in \mathcal{A}$ can be represented exactly as a
\emph{finite} polynomial series containing $N$ Kroneker
delta-symbols: a specific decomposed form of the CPF, which
expresses some ``independence'' of the random variables $a$ and
spatial coordinates $i$,
\begin{eqnarray} \label{Gen-P}
\!\!\!\!\!\!\!&&\!\!P(.|.)=P(a_i=\alpha|a_{i-N}^{i-1})
\\ [6pt] 
\!\!\!\!\!\!\!&& \!\!\!\!\!\!\!\!\!\!=\sum_{\beta_1... \beta_N \in
\mathcal{A}}\sum_{r_1...r_N} F_{\alpha;\beta_1\!\!...
\beta_N}(r_1,\!...,r_N) \!\!\prod_{s=0}^N
\delta(a_{i-r_s},\beta_s).\nonumber
\end{eqnarray}
Here the arguments $r_1,...,r_N$ of the function
$F_{\alpha;\beta_1... \beta_N}(r_1,...,r_N)$, supposed to be ordered
$r_1 < r_2 < ...< r_{N-1} < r_N$, indicate the distances between the
final ``generated'' symbol $a_i=\alpha$ and symbols $a_{i-1}, ... ,
a_{i-N}$. It is clear that there is one-to-one correspondence
between $P(a_i=\alpha|a_{i-N}^{i-1})$ and the function
$F_{\alpha;\beta_1...\beta_N}(r_1,...,r_N)$, which is referred to as
the \textit{generalized memory function}. The characteristic
functions $\delta(a_{i-r_s},\beta_s)$ play the pole of basis and the
generalized memory functions are coordinates of the CPF. We hope the
reader paid attention to the difference between $\alpha^k, k\in
(1,...,m)$ and $\alpha_s$, where subscript $s, s\in (1,...,N)$
enumerate different sets of letters.

It is helpful to decouple the memory function
$F_{\alpha;\beta_1...\beta_N}(r_1,...,r_N)$ and present it in the
form of the sum of \emph{memory functions} of $k$th \emph{order},
$F^{(k)}=F_{\alpha;\beta_{i_1}...
\beta_{i_k}}(r_{i_1},...,r_{i_k})$,
\begin{equation} \label{DecoupF}
F_{\alpha;\beta_1... \beta_N}(r_1,\!...,r_N) =\sum_{k=0}^N
F_{\alpha;\beta_{i_1}... \beta_{i_k}}(r_{i_1},...,r_{i_k}),
\end{equation}
where all symbols $r_{i}$ at the right hand side of
Eq.~\eqref{DecoupF} are different, ordered
\begin{equation} \label{DecoupF2}
 r_1 \leqslant r_{i_1} < r_{i_2}<...< r_{i_k}\leqslant r_N,
\end{equation}
and contain all different sets $\{r_{i_1},\!...,r_{i_k}\}$ picked
out from the set $\{r_1,...,r_N\}$. Inserting \eqref{DecoupF} in
\eqref{Gen-P} we get
\begin{equation} \label{PresTotP}
P(a_i=\alpha|a_{i-N}^{i-1}) = \sum_{k=0}^N Q^{(k)}(.|.)
\end{equation}
with
the following definitions of $Q^{(0)}$ and $Q^{(1)}$, see
Eqs.~(\ref{Dev_eq2}) and ~(\ref{Gen-P}),
\begin{equation} \label{P0}
Q^{(0)}= \!p_{\alpha},
\end{equation}
and
\begin{equation} \label{P0-Q1}
Q^{(1)}=  \sum_{\beta \in \mathcal{A}}\sum_{r=1}^N F_{\alpha
\beta}(r)[\delta(a_{i-r},\beta)\!-\!p_{\beta}].
\end{equation}
%

The general term of $k$th order, referred in multi-linear algebra to
as a $k$-linear form, is
\begin{eqnarray} \label{P-k}
\!\!\!\!\!\!\!\!&&\!\!Q^{(k)}(.|.)=\!\!\!\!\!\!\!\sum_{\beta_1...
\beta_k \in \mathcal{A}}\,\,\sum_{1\leqslant r_1 <...< r_k \leqslant
N}F_{\alpha;\beta_1... \beta_k}(r_1,\!...\!,r_k)  \\
[6pt] \!\!\!\!\!\!\!&&\!\!\!\!\!\times
\!\{\prod_{s=1}^k[\delta(a_{i-r_s},
\beta_{s}\!)\!-\!p_{\beta_{s}}\!]
-C_{\beta_k...\beta_1}\!(r_k\!-\!r_{k-1},\!...,r_k\!-\!r_1)\}.\nonumber
\end{eqnarray}

In Eq.~\eqref{P-k} we have added the term $C_{\beta_1...\beta_k}$ to
provide the equality $\overline{Q^{(k)}(.|.)}=0$ for all
$k=1,2,\ldots,N$.

Definition Eq.~\eqref{def_cor1} is correct for $r_i>0$,
$i=1,\ldots,k-1$. If some arguments of the correlation function in
Eq.~\eqref{def_cor1} are negative one should interpret it by the
following way, which is referred \cite{highord} to as the
\textit{''collating''}; we should order the arguments of the
function according to definition~\eqref{def_cor1}. For example,
\begin{eqnarray}\label{Coll}
&&\!\!\!\!\!C_{\alpha\beta\gamma\delta}(2,2,-3)=\\  [6pt] \nonumber
&&\!\!\!\!\!\overline{\big[\delta(a_0, \alpha)\!-\!p_{\alpha}
\big]\!\big[ \delta(a_{2}, \beta)\!-\!p_{\beta}\big]\big[\delta(a_4,
\gamma)\!-\!p_{\gamma} \big]\!\big[ \delta(a_{1},
\delta)\!-\!p_{\delta}\big]}\\ [6pt] \nonumber
&&\!\!\!\!\!=\!\!\overline{\big[\delta(a_0, \alpha)\!-\!p_{\alpha}
\big]\!\!\big[ \delta(a_{1}, \delta)\!-\!p_{\delta}\big]\!\!\big[
\delta(a_{2}, \beta)\!-\!p_{\beta}\big]\!\!\big[\delta(a_4,
\gamma)\!-\!p_{\gamma} \big]\!} \\ [6pt] \nonumber
&&\!\!\!\!\!=C_{\alpha\delta\beta\gamma}(1,1,2).\nonumber
\end{eqnarray}
We used this method to represent the correlation function in
Eq.~\eqref{P-k} in the collating form.

The function $C_{\alpha_1,...\alpha_k}$ depends on $k$ arguments,
but the Markov chain under consideration is supposed to be
homogenous. Then function $C_{\alpha_1,...\alpha_k}$ depends on
$k-1$ arguments, the differences between the indexes, $r_1,r_2,
\ldots, r_{k-1}$, of neighbor symbols. A trivial property of the
function $C_{\alpha_1,...\alpha_k} (r_1,\ldots,r_{k-1})$ is
\begin{eqnarray}\label{Ck-sym}
\sum_{\alpha_m \in \mathcal{A}} C_{\alpha_1,...\alpha_k}
(r_1,\ldots,r_{k-1}) = 0, \quad 1\leqslant m \leqslant k.
\end{eqnarray}
%


The last term $Q^{(N)}$ in Eq.~\eqref{PresTotP} contains all
different arguments $r_1,...,r_k$. For each fixed set of symbols
$\alpha;\beta_1... \beta_N$ there is just one function-constant
$F_{\alpha;\beta_1... \beta_N}(1,...,N)$.

The \emph{k-linear} conditional probability function $P^{(k)}(.|.)$
in the form of Eqs.~\eqref{PresTotP}, ~\eqref{P-k} can reproduce
correctly the correlations of the Markov chain up to the $k$th
order. For the value of $k=N$ the function $P^{(N)}(.|.)$ represents
exactly the function $P(a_i=\alpha|a_{i-N}^{i-1})$,
Eq.~(\ref{soglas}).

Thus, the conditional probability function
$P(a_i=\alpha^k|a_{i-N}^{i-1})$ is presented as decomposition of
multi-liner form into $k$-linear subspaces. Hosseinia, Leb and
Zideka~\cite{Hoss} proved rigorously that the conditional
probability can be written as a linear combination of the monomials
of past process responses for the Markov chain. Earlier this idea
was presented in Besag's paper \cite{Besag}. 

The utility of the decomposition procedure can be explained in the
following way. First, the partial terms $Q^{(k)}$ of the CPF are
mainly responsible for reproducing the correlation properties of
$k$th order, see below Eq.~\eqref{Bil CPF} and second, can be
considered as asymptotic successive approximation for the CPF.

\section{Bilinear CPF}

The model of the additive high-order Markov chain is well
studied~\cite{RewUAMM}. In this Section we examine high-order Markov
chains with a \emph{bilinear} conditional probability function.

The right-hand side of Eq.~(\ref{Dev_eq2}) contains two first terms
of asymptotic expression of the exact form, Eq.~(\ref{Gen-P}). The
next term $Q^{(2)}$ is
\begin{eqnarray} \label{TwoMem}
\!\!\!\!\!\!\!\!&&Q^{(2)}(.|.)
=Q^{(2)}(a_i=\alpha|a_{i-N}^{i-1})\nonumber \\
\!\!\!\!\!\!\!&&= \sum_{\beta, \gamma \in
\mathcal{A}}\,\,\,\sum_{1\leqslant r_1 < r_2 \leqslant N}
F_{\alpha;\beta\gamma}(r_1,r_2)\times
\\ \!\!\!\!\!\!\!\!\!\!\!\!\!\!\!
&&\{[\delta(a_{i-r_1},\beta)-p_{\beta}][\delta(a_{i-r_2},\gamma)-p_{\gamma}]
-C_{\gamma\beta}(r_2-r_1)\}.\nonumber
\end{eqnarray}
The conditional probability function, which contains linear term
$P_{add}=P^{(1)}(.|.)$ and bilinear function $Q_2$ (see
Eqs.~(\ref{Dev_eq2}) and~(\ref{TwoMem})) defines the Markov chain
with the bilinear memory function,
\begin{equation} \label{P2}
P_{bilin}(.|.)= P^{(2)}(.|.)= P^{(1)}(.|.)+Q^{(2)}(.|.).
\end{equation}

It is possible to find the recurrence relations for the correlation
functions of the $N$-step bilinear Markov chain. For this purpose
first of all we should calculate explicitly the average value of
symbol $a_{r_1+\ldots+r_{k-1}}$ in Eq.~\eqref{def_cor1}. Taking into
account the equation $P(a=\alpha|\cdot)+ P(a\neq \alpha|\cdot)=1$ we
can rewrite Eq.~\eqref{def_cor1} for arbitrary $k\geqslant 2$ in the
form,
\begin{equation}\label{Calc-Ck}
\overline{ [\delta(a_{0},\beta_1)-p_{\beta_1}]\ldots
[\delta(a_{r_1+\ldots+r_{k-1}},\beta_k)-p_{\beta_k}]}
\end{equation}
\begin{equation}
\!\!\!\!=\overline{[\delta(a_{0},\beta_1)-p_{\beta_1}]\ldots
[P(a_{r_1+\ldots+r_{k-1}}=\beta_k|\cdot)- p_{\beta_k}] }, \nonumber
\end{equation}
where CPF $P(a_{r_1+\ldots+r_{k-1}}=\alpha_k|\cdot)$ is given by
Eq.~\eqref{Gen-P}. In that way, we can obtain the fundamental
recurrence relation connecting the correlation functions of
different orders~$k$. Here we restrict ourselves by presenting these
equations for the correlation functions $C_{\alpha\beta}(r)$ and
$C_{\alpha\beta\gamma}(r_1,r_2)$ and bilinear CPF, Eq.~(\ref{P2}),

\begin{eqnarray} \label{BilinCorr2}
&&C_{\alpha\beta}(r)= \sum_{r_1=1}^{N} \sum_{\gamma \in
\mathcal{A}}C_{\alpha \gamma}(r-r_1) F_{\beta \gamma}(r_1) \\ [6pt]
\nonumber \!\!\!\!\!\!\!&&+
 \sum_{\gamma, \varepsilon\in \mathcal{A}}\sum_{1\leqslant r_1 < r_2 \leqslant
N}
C_{\alpha\varepsilon\gamma}(r-r_2,r_2-r_1)F_{\beta;\gamma\varepsilon}(r_1,r_2).
\end{eqnarray}
Equation for $C_{\alpha\beta\gamma}(r_1,r_2)$ reads
\begin{eqnarray} \label{BilinCorr3}
&&C_{\alpha\beta\gamma}(r_1,r_2)= \sum_{\eta \in
\mathcal{A}}\sum_{r'_1=1}^{N} C_{\alpha\beta\eta}(r_1,r_2-r'_1)
F_{\gamma \eta}(r'_1) \nonumber
\\ [4pt]
&&+ \sum_{\eta, \varepsilon\in \mathcal{A}} \sum_{1\leqslant r'_1 <
r'_2 \leqslant N} F_{\gamma;\eta\varepsilon}(r'_1,r'_2) \times\\
[6pt] \nonumber &&[C_{\alpha\beta\varepsilon\eta}
(r_1,r_2\!-\!r'_2,r'_2\!-\!r'_1)
-C_{\alpha\beta}(r_1)C_{\varepsilon\eta}(r'_2-r'_1)]. \nonumber
\end{eqnarray}

The other way to get Eqs.~(\ref{BilinCorr2}) and~(\ref{BilinCorr3})
is based on the minimization of the ``distance'',
Eq.~(\ref{Dev_eq40}).


The system of equations (\ref{BilinCorr3}) allows us to find the
unknown memory functions $F_{\alpha\beta}(r)$ and
$F_{\alpha\beta\gamma}(r_1,r_2)$ for consecutive constructing a
representative random sequence with given correlation functions of
the second and third orders. The memory functions should be
expressed by means of the probability $p_{\alpha}$ and the
correlation functions $C_{\alpha\beta}(r)$ and
$C_{\alpha\beta\gamma}(r_1,r_2)$, which can be found numerically by
means of analysis of a given random chain.

Equations~(\ref{BilinCorr2}) and~(\ref{BilinCorr3})
enable us to understand, that higher-order correlators and all
correlation properties of higher orders are not independent anymore.
We cannot control them and reproduce correctly by means of the
memory function $F_{\alpha \beta}(r)$ because the latter is
completely determined by the pair correlation function. Really, when
we generate an additive Markov chain, the second order memory
function $F_{\gamma;\eta\varepsilon}(r'_1,r'_2)$ equals zero.
Correlation function of the third order
$C_{\alpha\beta\eta}(r_1,r_2)$, determined by
Eq.~(\ref{BilinCorr3}), depends on $F_{\beta \gamma}(r_1)$.

We can make a similar conclusion about $k$th order memory functions.
They generate sequences for which we control the  $k$th order
correlation function. The $N$ step Markov chain with $k$-linear
memory function allows us to reproduce correctly the chains up to
the correlation function of $k$th order.

\section{Approximate solution of equations} 

Equations~(\ref{BilinCorr2}) and~(\ref{BilinCorr3}) can be
analytically solved only in some particular cases: for one- or
two-step chains, the Markov chain with a step-wise memory function
and so on. Here we give their approximate solution supposing that
correlations in the sequence are not too strong (in amplitude, but
not in length) and the alphabet $\mathcal{A}$ contains many letters.
In order to formulate these conditions we introduce the
\emph{normalized} symbolic correlation function defined by
\begin{equation}\label{Def K}
K_{\alpha \beta}(r)=\frac{C_{\alpha \beta}(r)}{C_{\alpha \beta}(0)},
\quad C_{\alpha \beta}(0)=p_{\alpha}\delta(\alpha,\beta)-p_{\alpha}
p_{\beta}.
\end{equation}

If correlations in the random chain are not strong, it is plausible
to suppose that the all components of the normalized correlation
function with $r \neq 0$ are small with respect to $K_{\alpha
\beta}(0)=1$.

Neglecting the second term in the right-hand side of
Eq.~(\ref{BilinCorr2}) (correctness of this approximation is
explained below, after Eq.~(\ref{Change-k-lin})) we get
Eq.~(\ref{Dev_eq4}). The solution of this equation can be written in
the form
\begin{equation}\label{FAdditive}
 F_{\alpha\beta}(r) = \frac{1}{p_{\beta}}C_{\beta\alpha}(r)
\end{equation}
if in definition \eqref{Def K} of $C_{\alpha \beta}(0)$ we can
neglect the term $p_{\alpha} p_{\beta}$ with respect to
$p_{\alpha}$. This is possible if the dimension $|\mathcal{A}|$ of
alphabet $\mathcal{A}$ satisfies the condition
\begin{equation}\label{DimA}
 |\mathcal{A}|=m \gg 1,
\end{equation}
so that all probabilities $p_{\alpha}$ are small.

It is easy to see that after substituting Eq.~\eqref{FAdditive} into
~\eqref{Dev_eq2}, we can rewrite the additive conditional
probability in the intuitively clear form,
%
\begin{eqnarray} \label{P_add_weak2}
P^{(1)}(a_i=\alpha|a_{i-N}^{i-1}) = p_\alpha + \sum_{r=1}^N \left[
P(a_i=\alpha|a_{i-r}) - p_\alpha \right],
\end{eqnarray}
which explains probabilistic meaning of Eq.~(\ref{FAdditive}) -- in
this approximation each symbol $a_{i-r},\, 1\leqslant r \leqslant N
$ effects independently on the probability to generate  $a_i$.

Our analysis of Eq.~(\ref{BilinCorr3}) shows that we can neglect the
first term in RHS of Eq.~(\ref{BilinCorr3}) with respect to the term
$C_{\alpha\beta\gamma}(r_1,r_2)$ in LHS because $F_{\gamma
\eta}(r'_1)$ contains only nondiagonal small component of
$C_{\eta\gamma}(r'_1)$). By the same reason the term
$C_{\alpha\beta}(r_1)C_{\varepsilon\eta}(r'_2-r'_1)$ is small with
respect to $C_{\alpha\beta\varepsilon\eta}
(r_1,r_2-r'_2,r'_2-r'_1)$. The last statement follows from
estimation of the correlator $C_{\alpha\beta\varepsilon\eta}
(r_1,r_2-r'_2,r'_2-r'_1)$. Its largest unique component satisfying
the conditions $r_2'\geqslant r_1 '+1, r'_1 \geqslant 1$ is
$C_{\alpha\beta\alpha\beta} (r_1,-r_1,r_1)$ at
$r'_1=r_2,r'_2=r_1+r_2$. Thus, Eq.~(\ref{BilinCorr3}) reduces to
\begin{eqnarray} \label{FBilin}
\!\!\!\!C_{\alpha\beta\gamma}(r_1,r_2)=F_{\gamma;\alpha\beta}
(r_2,r_1+r_2)C_{\alpha\beta\alpha\beta}(r_1,-r_1,r_1).
\end{eqnarray}

Taking into account Eq.~(\ref{DimA}) and neglecting correlations
while calculating $C_{\alpha\beta\alpha\beta} (r_1,-r_1,r_1)$ we get
\begin{eqnarray} \label{FBilin2}
\!\!\!\!\!\!\!\!\!F_{\gamma;\alpha\beta}(r_1,r_2)=
\frac{1}{p_{\alpha}p_{\beta}}C_{\alpha\beta\gamma}(r_2-r_1,r_1) .
\end{eqnarray}

Equation~(\ref{Gen-P}) for the conditional probability function of
the symbolic high-order Markov chain with bilinear memory function
(in the first approximation with respect to the small parameters
$|C_{\alpha \beta}(r)| \ll 1, \, r\neq 0$ and a multi-letter
alphabet, $p_{\alpha}\ll1$\,) takes the form
%
%
\begin{eqnarray}\label{Bil CPF}
&&P^{(2)}(a_i=\alpha|a_{i-N}^{i-1})\simeq p_{\alpha} \\ \nonumber
&&+ \sum_{\beta \in A}\sum_{r_1=1}^N
\frac{1}{p_{\beta}}C_{\beta\alpha}(r_1)[\delta(a_{i-r_1},\beta)-p_{\beta}]\\
\nonumber &&+
 \sum_{\beta, \gamma \in \mathcal{A}}\sum_{1\leqslant r_1 < r_2 \leqslant
N} \frac{1}{p_{\beta}p_{\gamma}}
C_{\beta\gamma\alpha}(r_2-r_1,r_1)\times
\\ \!\!\!\!\!\!\!\!\!\!\!\!\!\!\!
&&\{[\delta(a_{i-r_1},\beta)-p_{\beta}][\delta(a_{i-r_2},\gamma)-p_{\gamma}]
-C_{\gamma\beta}(r_2-r_1)\}.\nonumber
\end{eqnarray}
%

\section{K-linear form of the CPF}

Equations (\ref{FAdditive}) and (\ref{FBilin2}) show that we can
hope to obtain the similar expressions for generalized memory
functions of $k$th order expressed by means of correlation function.
Really, the result of such calculations can be presented in the form
%
\begin{eqnarray} \label{Change-k-lin}
&&\!\!\!\!\!\!F_{\alpha;\alpha_{1}...\alpha_{k}}
(r_1,...,r_k) \\ [6pt]
&&\!\!\!\!\!\! = \frac{1}{p_{\alpha_1}...p_{\alpha_k}}
C_{\alpha_{k}...\alpha_{1}\alpha}(r_k-r_{k-1},...,r_2-r_1,r_1).  \nonumber
\end{eqnarray}
To obtain this result let us summarize the main steps of this
procedure:
\newline (i) calculate the correlation function of $(k+1)$th order,
$C_{\alpha_{1}...\alpha_{k+1}}(r_1,...,r_k)$, $1\leqslant k\leqslant N$;
\newline (ii) while calculating use $P(.|.)= P^{(N)}(.|.)= \sum_{k=0}^N Q^{(k)}(.|.)$.
\newline (iii) as a result the correlation function is presented as a sum
of the memory functions of different order from 1 to $N$ with
coefficients $C_{\alpha'_{1}\alpha'_{2}...}(r'_1,r'_k,...)$.
\newline (iv) in the main term of the sum containing $\sum_{r',\beta }
F_{\alpha_{k+1};\beta_{1}...\beta_{k}} (r'_1,...,r'_k)\times
C_{\alpha_1...\alpha_k \beta_k ...\beta_1} (r_1,...,r_{k-1},
r_k-r'_k, r'_{k-1}-r'_{k-2},r'_2-r'_1)$ find the maximal term
$C_{\alpha_1...\beta_1} (r_1,...,r'_2\!-\!r'_1)$. It is maximal if
in two increasing sequences $0,r_1,\!...\!,r_{k-1}$ and $r_k\!-
\!r'_k, r'_{k-1}\!-\!r'_{k-2},r'_2\!-\!r'_1$ (rewritten accordingly
to the collating procedure as $0,r_1,r_1+r_2, ...,
r_1+r_2+...+r_{k-1}$ and $r_1+...+r_k-r'_{k},  r_1+...+r_k-r'_{k-1},
..., r_1+...+r_k-r'_{1}$)
 there is one-to-one correspondence among their terms: $r'_{1}=r_k,
 r'_2=r_k+r_{k-1}$ and $r'_k=r_k+...+r_{1}$ .
\newline (iv) neglact correlations while obtaing
$C_{\alpha_1...\beta_1}(r_1,...,r'_2\!-\!r'_1)=\prod_{s=1}^k p_{\alpha_s}
\delta(\alpha_s,\beta_s)$;
\newline (v) all others terms containing $F_{\alpha_{s+1};\beta_{1}...\beta_{s}}
(r'_1,...,r'_s)$ with $s\neq k$ are small with respect to that taken into account;
they contain additional small factors $p^r, r\geqslant 1$;

Thus, the conditional probability function Eq.~(\ref{Gen-P}) for the
symbolic high-order Markov chain in the first approximation with
respect to the small parameters $|C_{\alpha \beta}(r)| \ll 1, \,
r\neq 0$ and a multi-letter alphabet, $p_{\alpha}\ll 1$\, is
expressed by means of ``experimentally'' measured quantities, i.e.,
the correlation functions. Taking into account
property~(\ref{Ck-sym}), $\sum_{\alpha_m}
C_{\alpha_1,...\alpha_k}(r_1,\ldots,r_{k-1}) = 0$, it is convenient
to present the final main result of the paper in the form of series
Eq.~(\ref{PresTotP}), where we should substitute Eq.~(\ref{P-k}) and
replace the memory function by Eq.~(\ref{Change-k-lin}),
\begin{eqnarray} \label{Bil CPF_Short}
\!\!\!&&P(a_i=\alpha|a_{i-N}^{i-1}) = \sum_{k =0}^N \sum_{\beta_1...
\beta_k \in \mathcal{A}}\sum_{1\leqslant r_1 <...< r_k \leqslant N}
\nonumber \\ [6pt] &&\prod_{s=1}^k p^{-1}_{\beta_{s}}
C_{\beta_{k}\!\!..\beta_{1}
\alpha}(r_k\!-\!r_{k-1},\!.. ,r_2\!-\!r_1,r_1)\times\\
\!\!\!\!\!\!&&\{\prod_{t=1}^k[\delta(a_{i-r_t},
\beta_{t}\!)\!-\!p_{\beta_{t}}\!]
-C_{\beta_k...\beta_1}\!(r_k\!-\!r_{k-1},\!...,r_k\!-\!r_1)\}.\nonumber
\end{eqnarray}
Equation~(\ref{Bil CPF_Short}) provides a tool for constructing weak
correlated sequences with given, prescribed correlation functions.
Note that the $i$-independence of the function
$P(a_i=\alpha|a_{i-N}^{i-1})$ provides homogeneity and stationarity
of the sequence under consideration. According to the Markov theorem
(see, e.g., Ref.~\cite{shir}), the finiteness of $N$ together with
the strict inequalities
\begin{equation}\label{ergo_m}
 0 < \!P(a_{i+N}\!=\alpha|a_{i}^{i+N-1})\! < 1, \, i \in \mathbb{N}_{+} = \{0,1,2...\}
\end{equation}
provides ergodicity of the random sequence.

We see that if the correlations are week, all terms of the CPF are
independent of each other. If, e.g., we generate a sequence using
the terms of zero order and bilinear one we find that all
correlators are equal to zero except the third order correlator.  In
the general case it is not correct. When, e.g., we generate a
sequence with an additive memory function there appear correlations
of all order, not only pair ones.

\section{What to do if $p_{\alpha}$ are not small}

If the real sequence under study does not satisfy the condition
$p_{\alpha}\ll 1$, as, for example, it is for nucleotide DNA
sequences, where all four probabilities $p$ of the different
nucleotide occurring is of the order $1/4$, we cannot apply
Eq.~(\ref{Bil CPF_Short}) for obtaining the CPF by means of
correlation functions. To make this possible we should decrease the
probabilities $p$. For this purpose we could use the idea proposed
by Jim\'enez-Monta\~no, Ebeling, and others \cite{Jim,Jim2,Rapp} who
 suggested coding
schemes of the non-sequential recursive pair substitution. Each
successive substitution is accompanied by decrease of probability
$p_{\widetilde{\alpha}}$, where $\widetilde{\alpha}$ belong to a new
extended alphabet.

%

\section{Artificial neural network}\label{ANN}

Above we presented the analytical method of finding the memory
functions $F_{\alpha;\beta_1... \beta_N}(r_1,...,r_N)$ and expressed
them in terms of correlation functions. In this section, we expose
briefly numerical method of network training -- estimation of
unknown parameters in a network. The result of this procedure should
be the values of the matrix-functions $F(.)$.

According to definition \cite{Haykin}, artificial neural networks
(ANN) are a family of connectionist models used to estimate or
approximate functions (in our case it is
$P(a_i=\alpha|a_{i-N}^{i-1})$) that can depend on a large number of
generally unknown parameters. Artificial neural networks are
generally presented as systems of interconnected nodes or
``neurons''. The connections have numeric weights that can be tuned
based on experience, making neural nets adaptive to inputs and
capable of learning. These  definitions correspond to our goal to
 numerically estimating the unknown functions-matrices
$F_{\alpha;\beta_1... \beta_N}(r_1,...,r_N)$ depending on a great
number of parameters, say $N\sim 10^5$ or more.

We mentioned that the equations for  the memory functions
$F_{\alpha;\beta_1... \beta_N}(r_1,...,r_N)$ can be obtained
analytically by means of minimization of the distance
(\ref{Dev_eq40}) (known in the ANN theory under the names of cost
function or average system error) between desired and actual neuron
output values, the elements of real referent sequence and the CPF.
The same distance can be used for purposes of numerical finding
unknown quantity -- generalized memory matrices $F(.)$ -- under the
network training.

The considered problem with (potentially) given random sequence
falls within the paradigm of supervised learning, which can be
thought of as learning with a ``teacher''. In supervised learning,
each example is a pair consisting of an input vector object,
$a_{i-N}^{i}$, and a desired output value $a_{i}=\alpha$ or, more
precisely, their conditional probabilities
$P(a_i=\alpha|a_{i-N}^{i-1}$). A supervised learning algorithm
analyzes the training data and produces an inferred function, which
can be used for mapping new examples. An optimal scenario will allow
for the algorithm to correctly determine the class of memory
functions.

A commonly used mean-squared error tries to minimize the average
squared error between the network's output and the target value over
all the example pairs. When one tries to minimize this cost using
gradient descent for the class of neural networks called
multilayered perceptrons, one obtains the common and well-known
backpropagation algorithm for training neural networks.




A number of supervised learning methods have been introduced in the
last two decade. In the Caruana and Niculescu-Mizil paper
\cite{Caruana} a reader can find a large-scale empirical comparison
between ten supervised learning methods: SVMs, neural nets, and so
on.



%

\section{Numerical simulations}

In this section, to verify our analytical results, 
we give examples of numerical generation of random sequences with
the state space of dimension $|\mathcal{A}|=2$ (the symbols-numbers
of the sequence can only take on the two values: $0$ or $1$). Let us
note, that for a binary sequence there is no distinction between
symbolic and numeric approaches: all symbolic correlation functions
can be expressed by means of numeric ones and vice versa. More
detailed explanation is outlined in Appendix.

It is supposed that the modeled statistical properties of the random
chain are determined by the probability of the symbols occurring,
$p_\alpha =1/2$, the additive part of the memory function:
\begin{equation}\label{Fr_Triangle}
F(r) = \left\{\begin{array}{l}
0.002 r,\qquad \qquad 1 \leq r \leq 5, \\[5pt]
0.02 - 0.002 r, \quad 5 < r \leq 10, \\[5pt]
0, \qquad \qquad \qquad 10 < r.
\end{array} \right.
\end{equation}
and  the exponential (with respect to the both arguments) bilinear
part of the memory function:
\begin{equation}\label{Fr12_Exp}
F(r_1,r_2) = 0.5 \exp(-0.5 r_1) \exp(-0.5 r_2),
\end{equation}
with the truncated parameter $N=20$, playing the role of the memory
length.

These memory functions allow us to calculate conditional probability
function~\eqref{Gen-P}, consisting in this case of three partes: the
zero-order function, $p_\alpha = 1/2$, the first-order additive
function $Q^{(1)}$, and the second order bilinear function
$Q^{(2)}$, where the last two terms are given by expressions
~\eqref{P0-Q1} and~\eqref{TwoMem}. Using $p_\alpha$ and
Eqs.~\eqref{Fr_Triangle} and~\eqref{Fr12_Exp}, it is possible to
build up four different CPFs. Taking, e.g., the zero-order function
only, we obtain uncorrelated sequence.

We construct numerically three different random sequences of length
$10^8$. The first one is generated by the additive probability
function $Q^{(1)}$, Eq.\eqref{Fr_Triangle}, the second sequence are
obtained with the bilinear part, Eq.\eqref{Fr12_Exp} (but without
additive part) and the third chain, the most general in our case, is
get by the CPF contained both terms~\eqref{Fr_Triangle} and
\eqref{Fr12_Exp}. All these CPFs contain, evidently, the zero-order
term, $p_\alpha$.

\begin{figure}[h!]
\center\includegraphics[width=0.44\textwidth]{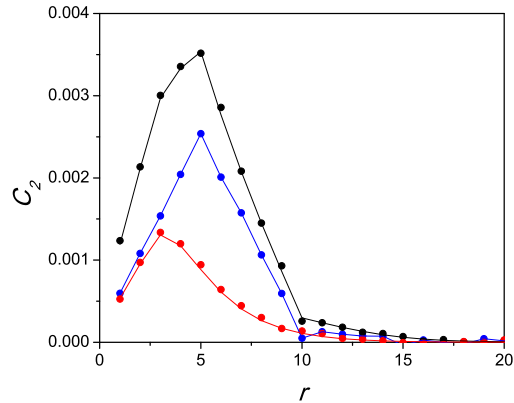}
\caption{(Color online) The pair correlation functions for three
random sequences constructed by means of the different conditional
probability functions taken in form of combination of the linear and
bilinear parts of the memory functions \eqref{Fr_Triangle} and
\eqref{Fr12_Exp}. The lines are the correlation functions obtained
by solving the system of equations~(\ref{BilinCorr2},
\ref{BilinCorr3}). The dots are direct calculations over generated
sequences of binary symbols. The length of sequences are $10^8$.}
\label{Fig_C2}
\end{figure}

Since the sequences are prepared, we can calculate their correlation
functions. In Fig.~\ref{Fig_C2}, the obtained correlators are
presented by the dots. At the same time, using  memory functions
Eqs.~\eqref{Fr_Triangle} and \eqref{Fr12_Exp} and solving
iteratively the system of equations Eqs.~(\ref{BilinCorr2},
\ref{BilinCorr3}) with respect to $C_{\alpha\beta}(r)$ and
$C_{\alpha\beta\gamma}(r_1,r_2)$, we obtain the correlation
functions presented in Fig.~\ref{Fig_C2} ($C_{\alpha\beta}(r) =
C_{11}(r)=C_2(r)$) by the curves.

The middle curve and dots correspond to the sequence generated with
the additive memory function. The bottom curve and dots describe a
sequence based on the bilinear part of the memory function. The
upper curve and dots present $C_2$ of the sequence obtained with the
additive and bilinear memory functions simultaneously.

From Eqs.~(\ref{BilinCorr2}, \ref{BilinCorr3}) and numerical
simulations, it follows that the correlation functions are entangled
and intricate. The additive part of the memory function is
responsible for not only second-order correlation function, the
bilinear part of the memory function effect on not only ternary
correlation function
---  they are mixed up in the system (of equations): each memory
function affects each correlator. The additive and bilinear parts of
the memory function is mainly responsible for for second-order  and
third-order correlation functions, correspondingly,  for a limiting
case of small correlations predominantly. Note, the ``small'' values
of the correlation functions $C_2$ of order of $10^{-3}$ are really
not small. We chosen the normalization coefficients of the memory
functions \eqref{Fr_Triangle} and \eqref{Fr12_Exp} in such way to
provide extremely strong persistent correlations with the  CPF
$P^{(2)}(a_i=\alpha|a_{i-N}^{i-1})$, Eq.~\eqref{P2}.

In the case of purely additive MF, the form of the correlator is
close to the shape of the memory function (middle curve-dots in
Fig.~\ref{Fig_C2}), but has a significant ``tail'' at distance $10
\lesssim r \lesssim 15$. This the phenomenon of ``lag'' of a
correlator with respect to a memory function, we have already seen
in previous studies ~\cite{muya}. In the absence of the additive MF
(bottom curve in Fig.~\ref{Fig_C2}), the pair additive correlator is
not equal to zero and has a shape defined by solution of
Eqs.~\eqref{BilinCorr2} and~\eqref{BilinCorr3}. In the case of the
simultaneous generation with additive and bilinear MFs, the pair
correlator becomes significantly different from the previous cases.

\begin{figure}[h!]
\center
\includegraphics[width=0.35\textwidth, height=0.3\textwidth
]{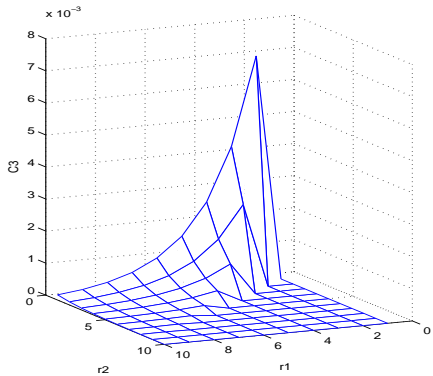} \caption{ The third-order correlation function $C(r_1,
r_2)$ of the random binary sequence constructed with using bilinear
exponential memory function ~\eqref{Fr12_Exp}.} \label{Fig_C3}
\end{figure}

\begin{figure}[h!]
\center
\includegraphics[width=0.44\textwidth]{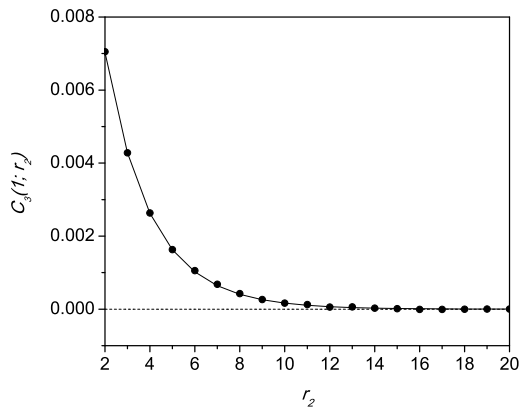} \caption{The
comparison of prescribed (line) and numerically found (dots) the
third-order correlation function $C(r_1, r_2)$ of the binary
sequence (built taking into account the additive~\eqref{Fr_Triangle}
and bilinear~\eqref{Fr12_Exp} memory functions) for the fixed
coordinate $r_1 = 1$. } \label{Fig_C3_1}
\end{figure}

The two dimensional surface for the third-order correlator $C(r_1,
r_2)$, 
obtained by solution of Eqs.~\eqref{BilinCorr2}
and~\eqref{BilinCorr3} with bilinear part of the memory function
Eq.~\eqref{Fr12_Exp} only is shown in Fig.~\ref{Fig_C3}. The
comparison of this function (line) and numerically found the
third-order correlation function $C(r_1, r_2)$ (dots) of the binary
sequence for the fixed coordinate $r_1 = 1$ is presented in
Fig.~\ref{Fig_C3_1}.

%
%

A good agreement between the analytical and numerical calculations
of the correlation functions in all three cases gives a reason to
believe that the system of equations and the generation procedure
based on the bilinear form of the CPF are well grounded.

\section{Conclusion}
The decomposition procedure for the CPF of symbolic random sequences
considered as the high-order Markov chains with long-range memory is
obtained. We represented the conditional probability function as the
sum of multi-linear memory function monomials of different orders.
It allows  us to build artificial sequences by method of successive
iterations taking into account at each step of iterations more and
more high correlations among random elements. At weak correlations,
the memory functions are uniquely expressed in terms of high-order
symbolic correlation functions. So we have filled the theoretical
gap between the methods of additive Markov chain and likelihood
estimation. The obtained results might be used for sequential
approximation of artificial neural networks training.


\section{Appendix}
Binary random sequences always can be considered as numeric and by
this reason equations~\eqref{BilinCorr2} and ~\eqref{BilinCorr3} can
be considerably simplified to a numeric form. Let us consider
symbolic correlator~\eqref{def_cor1} taken at $\alpha_1 = \alpha_2 =
\ldots = \alpha_k = 1$. In this case it is possible to put
$\delta(a_r,\alpha)=a_r$, whereupon the symbolic correlation
function takes on the form identical to the numeric one,
\begin{eqnarray}\label{C_symb_num_1}
  C_{1,...1}(r_1,r_2,\ldots,r_{k-1})&  =&\\
  \overline{(a_0 - p_1) \ldots (a_{r_1+\ldots+r_{k-1}} - p_1)}
  &\equiv & C(r_1,r_2,\ldots,r_{k-1}).\nonumber
\end{eqnarray}
To calculate a symbolic correlator with arbitrary indexes it is
sufficient to note that the term $\delta(a_r,\alpha) - p_\alpha$
changes its sign when one replaces $1$ to $0$ or  $0$ to $1$,
$\delta(a_r,\alpha) - p_\alpha = (1 - a_r) - p_0 = - (a_r - p_1)$.
Each such replacement changes a sign of the correlator.
Consequently, its value depends on the number $q$ of symbols ``0''
among $\alpha_1,...\alpha_k$,
\begin{equation}\label{C_symb_num}
    C_{\alpha_1,...\alpha_k}(r_1,r_2,\ldots,r_{k-1}) = (-1)^q
    C(r_1,r_2,\ldots,r_{k-1}).
\end{equation}
Using the same reasoning, we find a simplified expression for
$\delta(a_r,\alpha) - p_\alpha$ and the CPF,
\begin{equation} \label{P_symb_bin}
P(a_i = 1 | \ldots) = p_1 + \sum_{r=1}^N F(r)(a_{i-r} - p_1),
\end{equation}
where
\begin{equation} \label{F_symb_bin_2}
F(r) = F_{11}(r) - F_{10}(r).
\end{equation}
Let us rewrite Eq.~\eqref{BilinCorr2} for the element $C_{11}(r)$.
We put $\alpha=\beta=1$ and take a sum over $\gamma = \{0, 1\}$.
Then, at the LHS we obtain the numeric correlator $C(r)$; at the
RHS, we express all symbolic correlators by means of numeric ones,
Eq.~\eqref{C_symb_num}. After that, all symbolic memory functions
turn out to be grouped in combinations, which give numeric memory
functions, e.g.,
\begin{eqnarray}
\nonumber
C_{10}(r-r_1)F_{1;0}(r_1) + C_{11}(r-r_1)F_{1;1}(r_1) = \\
C(r-r_1)(F_{1;1}(r_1) - F_{1;0}(r_1)) = C(r-r_1)F(r_1).
\end{eqnarray}
In the same way we manipulate with the second term in
Eq.~\eqref{BilinCorr2} and introduce the second order numeric memory
function
\begin{eqnarray}\label{F_symb_bin_3}
\nonumber
  F(r_1, r_2) &=& F_{100}(r_1, r_2) - F_{101}(r_1, r_2) \\
  &-& F_{110}(r_1, r_2) + F_{111}(r_1, r_2).
\end{eqnarray}
As a result, equations~\eqref{BilinCorr2} and ~\eqref{BilinCorr3}
take on the following simplified formes convenient for describing
binary sequences:
\begin{eqnarray} \label{BilinCorr2_num}
&&C(r)= \sum_{r_1=1}^{N} C(r-r_1) F(r_1) \\ [6pt] \nonumber &&+
\sum_{r_1=1}^{N-1}\sum_{r_2=r_1+1}^N C(r-r_2,r_2-r_1)F(r_1,r_2),
\end{eqnarray}
\begin{eqnarray} \label{BilinCorr3_num}
&&C(r_1,r_2) = \sum_{r'_1=1}^{N} C(r_1,r_2-r'_1) F(r'_1) \\
\nonumber &&+ \sum_{r'_1=1}^{N-1}\sum_{r_2'=r_1'+1}^N F(r'_1,r'_2)[C(r_1,r_2\!-\!r'_2,r'_2\!-\!r'_1) \\
\nonumber &&-C(r_1)C(r'_2-r'_1)].
\end{eqnarray}

\end{document}